\documentclass[a4paper]{jpconf}
\usepackage{graphicx}
\def\be{\begin{eqnarray}}
\def\ee{\end{eqnarray}}
\def\ba{\begin{array}}
\def\ea{\end{array}}

\def\hbb{\hbar}
\def\omc{\omega_c}

\begin{document}
\title{Distribution of an Ohmic current in the close vicinity of a quantum point contact}

\author{Afif Siddiki}

\address{Physics
Department, Arnold Sommerfeld Center for Theoretical Physics, and
CeNS, Ludwig-Maximilians-Universit\"at M\"unchen, Theresienstr.
37, D-80333 M\"unchen, Germany}

\ead{siddiki@lmu.de}

\begin{abstract}
We present the essential findings of the screening theory of the
integer quantum Hall effect (IQHE) considering a quantum point
contact (QPC). Our approach is to solve the Poisson and the
Schr\"odinger equations self-consistently, taking into account
electron interactions, within a Hartree type approximation for a
two dimensional electron gas (2DEG) subject to high perpendicular
magnetic fields. The Coulomb interaction between the electrons
separates 2DEG into two co-existing regions, namely quasi-metallic
compressible and quasi-insulating incompressible regions, which
exhibit peculiar screening and transport properties. In the
presence of an external current, we show that this current is
confined into the incompressible regions where the drift velocity
is finite. In particular, we investigate the distribution of these
incompressible strips and their relation with the quantum Hall
plateaus considering a quasi 1D constriction, i.e. a QPC.
\end{abstract}

\section{Introduction}
\begin{figure}
{\centering %
\includegraphics[width=.5\linewidth]{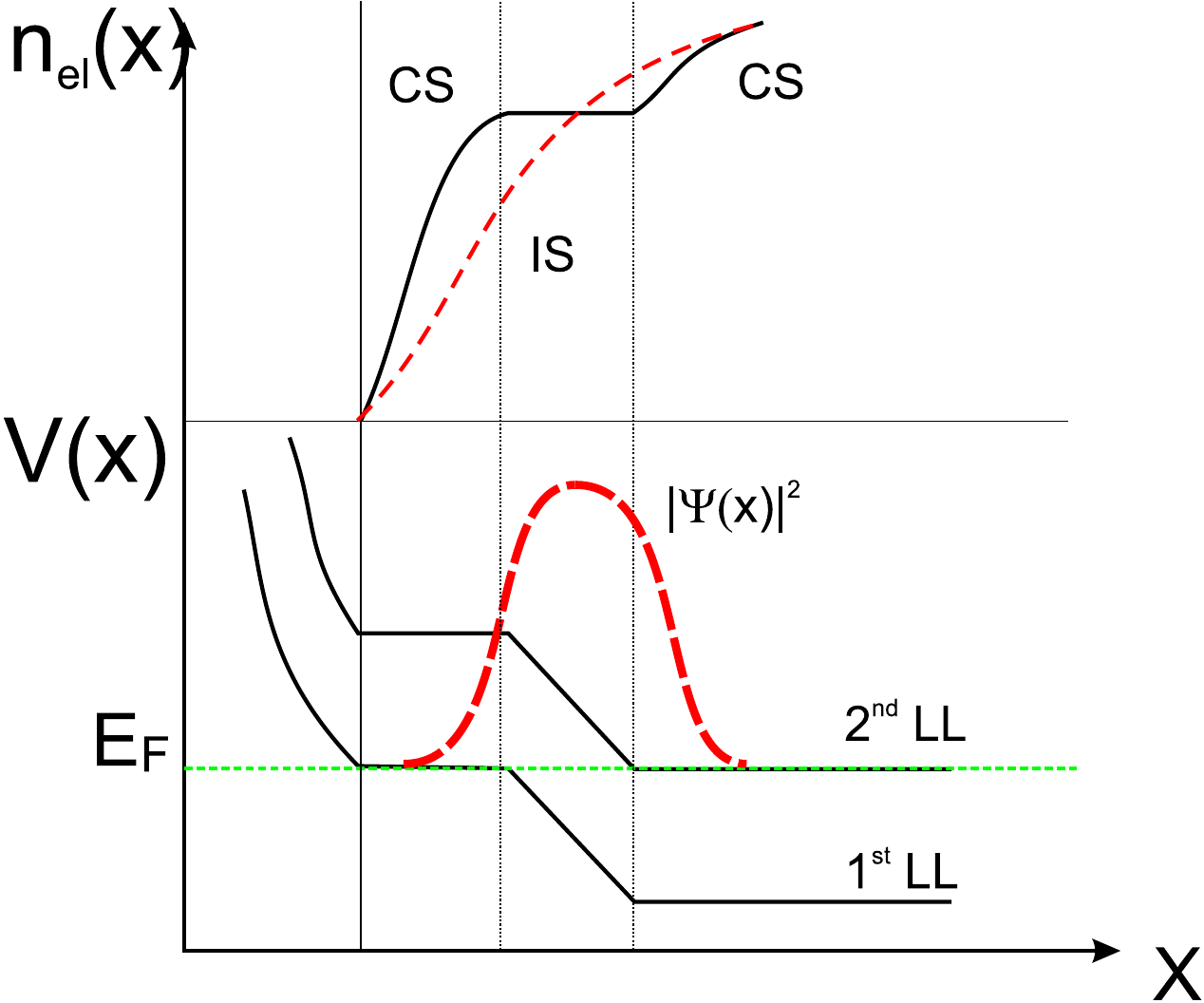}
\caption{ \label{fig:fig1} A sketch of the density profile as a
function of lateral coordinate (upper panel) and the corresponding
energy dispersion, i.e. 1$^{st}$ and 2$^{nd}$ Landau levels (lower
panel). The vertical dashed lines indicate the region where the
incompressible strip (IS) resides, whereas the horizontal line in
the lower panel is the Fermi energy. The dashed curve in the upper
panel is the density distribution, when one considers finite
extend of the wave function (dashed (green) curve in the lower
panel) and the corresponding curve calculated within the TFA.}}
\end{figure}

The increasing amount of interest in quantum information
processing attracted many experimentalists and theoreticians to
investigate the transport properties of low dimensional charge
systems. Beyond interest to build quantum computational units,
i.e. q-bits, and the calculation algorithms which are supposed to
be used by these units, the coherency of the information
processing, or in other words information transport, is an
essential challenge. Since some of the proposed units are defined
on the 2DEG, the information carriers are naturally the electrons.
In the absence of an external magnetic field, the well-known and
well-defined Landauer channels~\cite{Landauer81:91} are the best
candidates for coherent transport. These channels are by
definition ballistic and 1D. They are formed due to the size
quantization of the system under investigation. A similar
non-interacting single particle picture commonly used to describe
the transport, also in the presence of an external magnetic field,
is the so-called Landauer-B\"uttiker edge states
(LB-ES)~\cite{Buettiker86:1761}. However, its relevance in
explaining some of the recent quantum Hall, in particular the
local probe, experiments is questionable, where the importance of
electron-electron ($e-e$) interaction is shown to be
dominant~\cite{Ahlswede01:562,Yacoby04:328,Jose07phye}. The
simplest way of including $e-e$ interactions is the so-called
electrostatic approximation (ESA) which is based on the Landau
quantization introduced by the external magnetic
field~\cite{Chklovskii92:4026}. In this approach, the 2DEG is
split into two regions depending on the value of the Fermi energy
with respect to Landau level energy, $E_n=\hbb\omc(n+1/2)$, where
$\omc=eB/m^*$ is the cyclotron frequency and $m^*$ is the
effective mass of the electron ($=0.067m_e$), with Landau index
$n$. If the Fermi energy equals one of the Landau energies, which
has a $2\pi l^2-$fold degeneracy, the electronic system is
quasi-metallic so-called compressible and screening is nearly
perfect. Otherwise, when the Fermi energy lies in between two
consequent Landau levels, screening is poor and system is called
to be incompressible. Beyond the differences in their screening
properties, the transport properties of compressible and
incompressible regions are different as black and white. First of
all we should mention that, within a compressible strip (CS) the
screened potential is (ideally) flat and density is varying, as a
result of high density of states (DOS) at the Fermi energy,
whereas within the incompressible strip (IS) the potential is
varying (since the external potential could not be screened
perfectly) and electron density is constant (see
Fig.~\ref{fig:fig1}). The discussion on "where the current flows
?" is a long lasting question. As early as, the formation of
CS/ISs was realized, it was conjectured that, the current should
flow from the IS, since the screened potential has a gradient only
within these regions~\cite{Chang90:871,Fogler94:1656} and back
scattering is
suppressed~\cite{Guven03:115327,siddiki2004,Akera06:}. However,
later it was argued that since there are no states available at
the Fermi energy the current should flow from the CSs, where it is
possible to generate an
excitation~\cite{Chklovskii92:4026,Beenakker89:2020}. Moreover,
the non-interacting limit of the CS was claimed to reduce to the
Landauer-B\"uttiker formalism~\cite{Beenakker89:2020}. The recent
local potential~\cite{Ahlswede02:165} and
transparency~\cite{Yacoby04:328} experiments supported the
findings of the first model, where it was shown that the positions
of the Hall potential drops coincide with the predicted positions
of the (innermost) ISs within ESA. Even before the ESA, a
self-consistent interaction model was developed by R.R. Gerhardts
and his co-workers~\cite{Wulf88:4218}, which excellently agrees
with the experimental results after including the influence of an
imposed current~\cite{Guven03:115327,siddiki2004}. In this model,
a Hartree type mean-field approximation was considered, which
provides the local distribution of the electron density and
potential profiles as a function of magnetic field and
temperature~\cite{Lier94:7757,Oh97:13519,Suzuki93:2986}. A local
transport model was incorporated to describe the current
distribution, where the entries of the conductivity tensor were
calculated from a reasonable conductivity model, e.g.
self-consistent Born approximation (SCBA)~\cite{Ando82:437},
provided that local electron distribution is known. Moreover, the
position dependent electrochemical potential, thereby the local
current distribution, was obtained within a local version of the
Ohm's law under the condition of a \emph{local
equilibrium}~\cite{Guven03:115327,velocitynldeniz07}. This
self-consistent theory was implemented to many interesting 2DEG
systems successfully explaining the relevant
experiments~\cite{siddikikraus:05,Bilayersiddiki06:,SiddikiMarquardt,deniz06,qdotphani07,qpcengin07}.

The recent quantum interference experiments considering 2DES under
strong magnetic fields, provided unexpected information on the
transport electrons. In the Mach-Zehnder interference
setups~\cite{Heiblum03:415,Neder06:016804,Litvin07:033315,Litvin:2008arXiv},
it was observed that the contrast in the interference oscillations
(visibility) is path-length independent, which is in strong
contradiction with the optical counterpart of the experimental
setup. The basic idea is to split a monochromatic (mono-energetic)
light (electron) beam from a half-transparent mirror (QPC) and
measure the interference after these two split beams interfere. In
these experiments the LB-ES are assumed to simulate the light beam
and QPCs to replace the mirrors. However, the out come of the
experiments were not easy to explain within a naive single
particle model, without many
assumptions~\cite{Samuelson04:02605,Neder07:112}. Another very
interesting experiment concerning QPCs is performed at the
Westervelt group (Harvard), where a local probe technique is used
to observe electron flow near a QPC~\cite{Tobias07:464}. There it
was clearly shown that the actual realization of the sample is
dominant in determining the transport taking place (see the paper
by T. Kramer within this issue). The essential physics was
explained comprehensively within the wave-packet formalism at
relatively low magnetic fields ($\sim 0.4$ T), where one should
not expect the formation of the incompressible strips. However, if
one considers higher magnetic fields one should in principle
include the effect of $e-e$ interactions in a self-consistent
manner. Beyond the interference experiments, the transmission
measurements performed at a 2DEG using a single QPC has shown
peculiar effects~\cite{Roddaro05:156804}. The suppression or
enhancement of the transmission amplitude presents significant
deviation from the chiral Luttinger liquid model~\cite{Dassarma},
meanwhile this effect could be explained by considering an
electron-hole asymmetry~\cite{Lal07:condmat}. In this treatment
the formation of a large incompressible region is assumed,
however, the $e-e$ interaction is not taken into account.

Here we provide, such a calculation scheme within Thomas-Fermi
approximation. In Sec.~\ref{sec:sec1} we briefly describe the
self-consistent model to obtain the electrostatic quantities and
discuss the limitations of this approach. We then proceed with the
formulation of the current flow within a local Ohm's law in Sec.
~\ref{sec:sec2}. We finalize, our work by showing the
interrelation between the quantized Hall plateaus (QHP) and the
expected distribution of the IS, and thereby local current, near a
QPC, also in the out of linear response regime.
\section{The electrostatic self-consistency\label{sec:sec1}}
\begin{figure}
{\centering
\includegraphics[width=.8\linewidth]{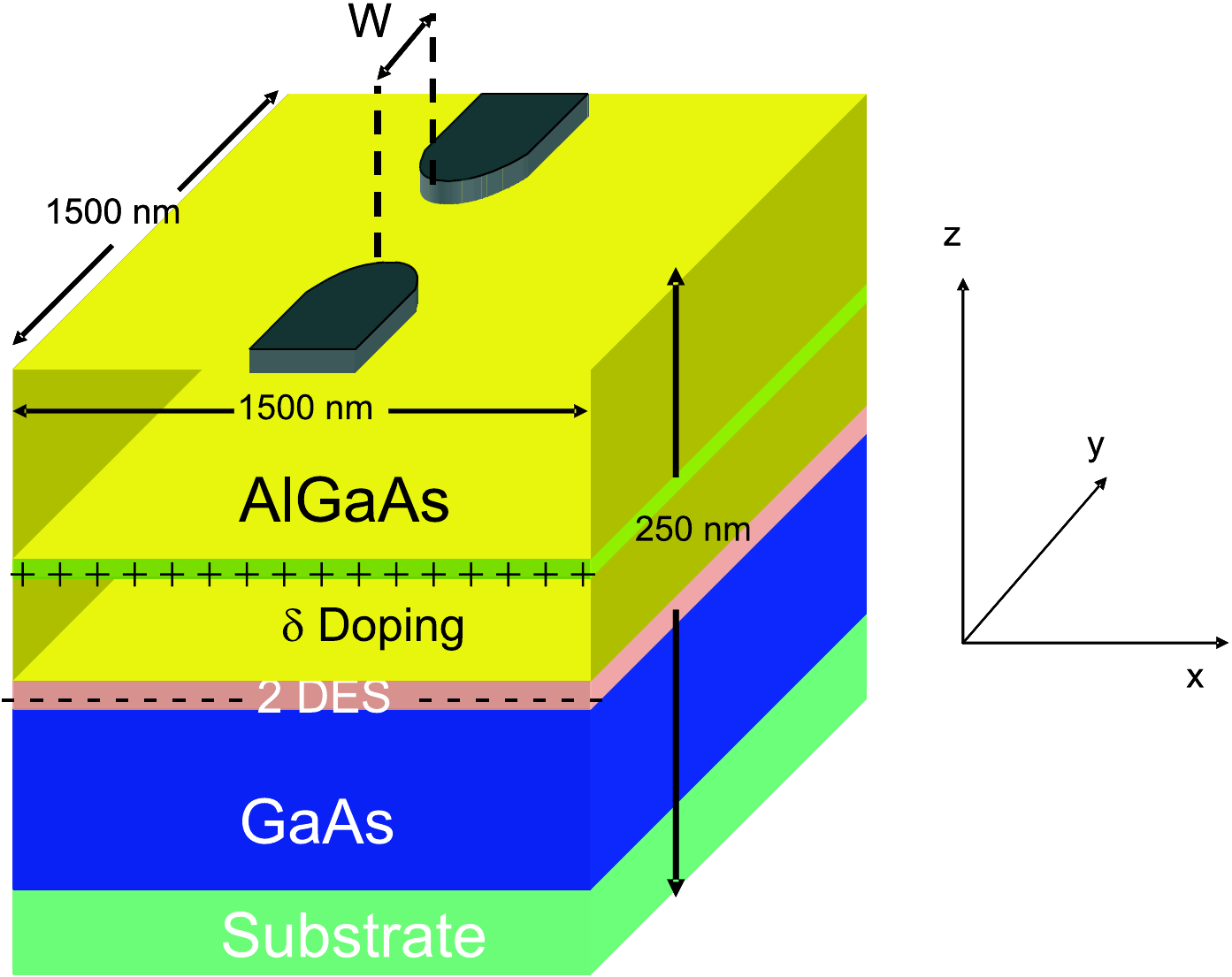}
\caption{ \label{fig:fig2}Schematic draw of the sample under
investigation. The gates defining the QPC are deposited on the
surface of the sample along $y-$ direction with a separation
$W\sim 150$ nm, under a potential bias $V_{\rm g}=-1.5$ V, such
that the 2DEG beneath is depleted. The width of the gates are
$\sim 500$ nm and assumed to be thinner than 10 nm. A homogeneous
donor layer with a number density $4\times10^{12}$ cm$^{-2}$ is
laying 43 nm below the surface, which donates almost 10 percent of
its excess electrons to the 2DEG. The resulting electron density
of the 2DEG is $3.0\times10^{11}$ cm$^{-2}$ without gates. In
order to satisfy the open boundary conditions a larger lattice is
considered in actual calculations spanned with
$128\times128\times32$ mesh points in $x$, $y$ and $z$ directions,
respectively.}}
\end{figure}

The essential physics of a 2DEG subject to strong perpendicular
magnetic field is governed by the single particle Hamiltonian \be
H=H_0+V(x,y) \label{eq:ham}\ee with the kinetic term \be
H_0=\frac{1}{2m^*}(\textbf{p}-\frac{e}{c}\textbf{A})^2
\label{eq:ho}\ee where $\textbf{p}$ stands for momentum and
$\textbf{A}$ for the vector potential. The potential energy is
composed of the external $V_{\rm ext}(x,y)$ and interaction
$V_{\rm H}(x,y)$ terms\be V(x,y)=V_{\rm ext}(x,y)+V_{\rm H}(x,y)
\label{eq:vext}.\ee Using a relevant gauge (here we use the Landau
gauge to exploit the symmetry of the system) and assuming a
translational invariance, one ends with the reduced 1D Hamiltonian
in $x-$ direction, which is nothing but a Harmonic oscillator
centered at $X=-l^2k_y$. The solutions of the Hamiltonian in $y-$
direction are properly normalized plane waves with
quasi-continuous momentum $k_y$ $(=\frac{n \pi}{L_y}$, where $L_y$
is the sample length), and $l^2=\hbb/m\omc$ is the square of the
magnetic length. By translational invariance we mean that, the
quantization effects arising from the finite length of the sample
are negligible. The solution of the Schr\"odinger equation in $x-$
direction results in the well-known Landau wave functions \be
\Phi_{n_X}(x)=\frac{1}{ \sqrt{2^n n! \sqrt{ \pi} l }}
\exp{[-(\frac{x-X}{l})^2/2]}\times H_n(\frac{x-X}{l}), \ee where
$n$ is the Landau index and $H_n(\xi)$ the $n$th order Hermite
polynomial with the argument $\xi$, whereas the eigen energies are
\be E_{n,X}=\hbar \omega_c(n+1/2).\ee The external potential is
obtained by solving the 3D Poisson
equation~\cite{Weichselbaum03:056707,Sefa08:} for the structure
shown in Fig.\ref{fig:fig2}a. The calculation scheme and its
details of implementing to a QPC is far beyond the scope of this
paper and is discussed elsewhere, however, the for the present
work a reasonable confinement potential profile is sufficient to
initialize the self-consistent scheme described below. The Hartree
potential is defined from the electron distribution via, \be
V_{\rm H}(x,y)=\frac{2e^2}{\bar{\kappa}}\int_{A}K(x,y,x',y')n_{\rm
el}(x',y')dx'dy',\label{eq:tfapotential}\ee where the kernel
$K(x,y,x',y')$ gives the solution of the Poisson equation at
$(x,y)$ for a point charge residing at the point $(x',y')$,
assuming periodic boundary conditions in 2D and $\bar{\kappa}$ is
an average dielectric constant defined by the sample parameters.
Once the Hamiltonian given in Eq.~\ref{eq:ham} is solved one can
obtain the electron density from, \be n_{\rm
el}(x,y)=\sum_{n,k_y}{|\psi_{n,k_y}(x,y)|^2}f(E_{n,k_y}-\mu^*),
\ee where f(E) is the Fermi function and $\mu^*$ is the (position
dependent, in the presence of an external current) electrochemical
potential.

We first make the assumption that, the total potential varies
smoothly on the scale of the magnetic length, i.e. Thomas-Fermi
approximation (TFA). This allows us to replace the wave functions
with delta functions, the eigen energies are then given by: \be
E_{n}(X,y)=E_n+V(X,y).\ee The above approximation results in a
simpler description of the electron density, \be n_{\rm
el}(x,y)=\int{dE f(E-\mu^*)D(E,(x,y))} \label{eq:tfaden}\ee within
the TFA, where $D(E,(x,y))$ stands for the local density of states
(LDOS). It is more common to use the filling factor $\nu(x,y)$
instead of the electron density, which is defined as
$\nu(x,y)=2\pi l^2 n_{\rm el}(x,y)$. The name itself is
explanatory, filling factor describes how many of the low-lying
Landau levels is filled. If it is a integer, it means that all the
states below the Fermi level are occupied (i.e. incompressible),
otherwise show how much of the top most Landau level has available
states.

The calculation procedure is as follows, we start with a given
external potential generated by the donors and gates, then
calculate the electron density distribution in the absence of an
external magnetic field. The average electron density,
$\overline{n_{\rm el}}=\int_A{n_{\rm el}(x,y)}$, is kept constant
all through the calculation, which enables to have a convergence
criteria. This electron density initializes the potential given in
Eq.\ref{eq:tfapotential}, which in turn determines the electron
density via Eq.\ref{eq:tfaden}. The numerical procedure is a bit
more complicated than described, however, the details can be found
in Ref.~\cite{SiddikiMarquardt,Sefa:Diss}.
Before proceeding with the discussion of the self-consistent
results obtained for a QPC geometry, we would like to make some
comments on the weak points of the TFA. The first issue is that,
if an IS becomes narrower than the magnetic length our
approximation fails~\cite{Suzuki93:2986,siddiki2004}. Such a
potential profile is shown in Fig.~\ref{fig:fig1}, therefore one
should avoid these situations and perform a spatial averaging over
the magnetic length both for the density and the potential
profiles as a first order approximation~\cite{siddiki2004}. This
approach has been used previously for these systems and is shown
to be sufficiently powerful to simulate the effect of the quantum
mechanical wave
functions~\cite{Siddiki04:condmat,Bilayersiddiki06:}. Another
issue about narrow ISs, is that the LDOS ($D(E,(x,y)$) become
broader, when considering steep potential drops, i.e. large
transverse electric fields~\cite{TobiasK06:h}. Due to this local
broadening of the LDOS the Landau gap is smeared out and as a
result IS disappears~\cite{Siddiki08:LDOS}. These two observations
together, i.e. overlap of the wave functions and local broadening,
justifies the validity of performing a spatial average over
magnetic length or even over the Fermi wavelength $\lambda_F$,
which we also will perform through out this work.

The sample that we investigate is shown in Fig.~\ref{fig:fig2}, a
multi-layer GaAs is grown on top of the substrate, which is
followed by a thick layer of AlGaAs. This layer is ($\delta$-)
doped by Silicon, which provide electrons to the 2DEG formed at
the interface of the two different materials. On top of the sample
(metallic) gates are patterned to manipulate the electron density
at the 2DEG. If a sufficiently large negative potential is applied
to these gates (here patterned as a QPC) the electrons underneath
the gates are depleted, however, the potential profile satisfies
the conditions of the TFA~\cite{Sefa08:,Sefa:Diss}. We will
consider a geometric pattern of the gates such that the opening of
the QPC is 150 nm and is smooth again on the length scales of the
magnetic length.
\begin{figure}
{\centering
\includegraphics[width=1.\linewidth]{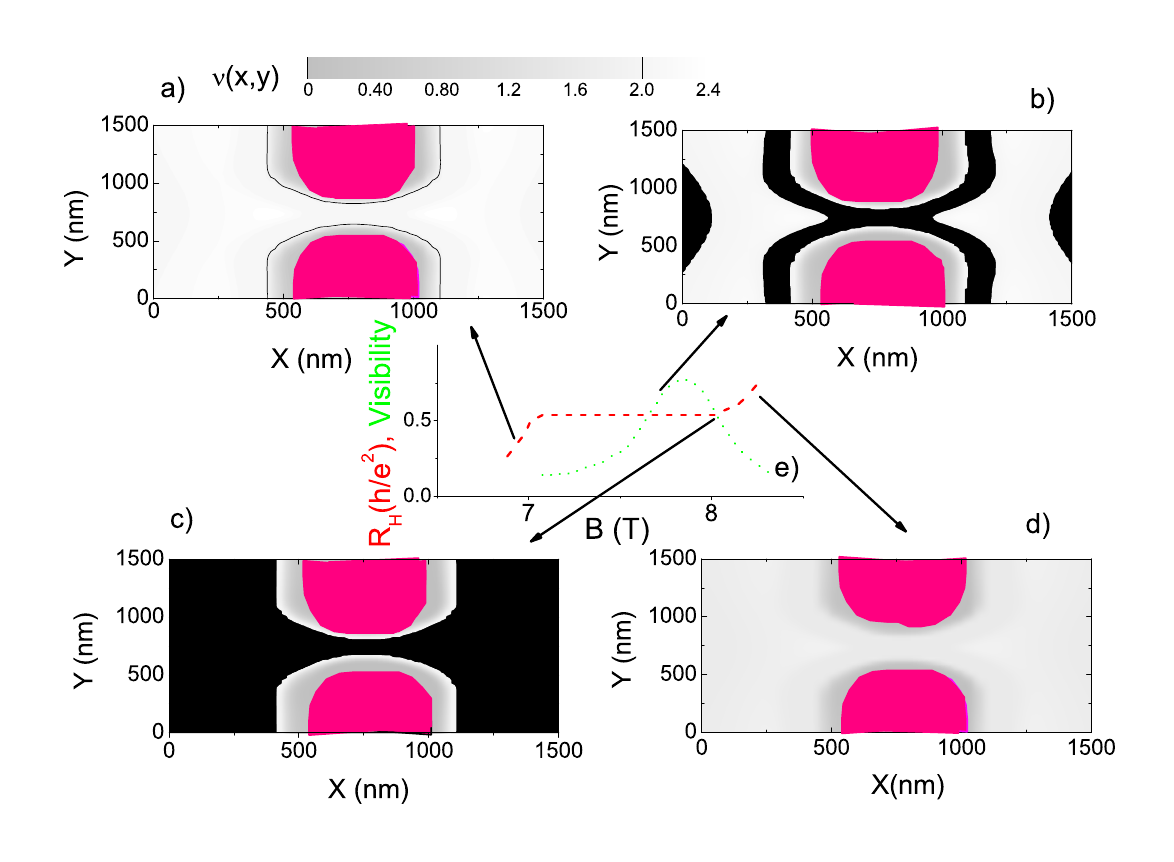}
\caption{ \label{fig:fig3} Spatial distribution of the electron
density (a-d) together with an illustrative sketch of the Hall
resistance (broken red line) and the expected visibility (dotted
green line) as a function of magnetic field strength (e). The
solid black regions indicate an integer filling factor equal to 2,
i.e. an IS, whereas the degrading gray scale denotes the
compressible regions with varying electron density, whereas the
colored (pink) region is the electron depleted region. The initial
potential calculated at $B=0$ is obtained for a geometry shown in
Fig.~\ref{fig:fig2}. The characteristic $B$ values are chosen
depending on the existence and the width of the ISs, such that the
resulting magnetic field strengths are $B=6.7$ T (a), $B=7.7$ T
(b), $B=7.9$ T (c), $B=8.3$ T (d). All the calculations are done
at temperatures satisfying $k_BT/\hbb\omc\ll 0.04$ for $I=0$. For
the DOS broadening parameter see the related text.}}
\end{figure}
Fig. ~\ref{fig:fig3} shows a sequence of density distributions
calculated at selected magnetic field values, for which the 2DEG
system is out of the $\nu=2$ QHP at low $B$ fields (a), at the
lower edge of the QHP (b) at the higher edge of the QHP (c) out of
the QHP at high $B$. Pedagogically, it is preferable to start the
discussion considering high magnetic fields since the density
distributions at $B=0$ and at strong $B$ are similar. At zero
magnetic field a 2DEG is known to be quasi-metallic, due to high
(but not infinite) DOS, screening is perfect and the external
potential is almost flat in the interior of the sample. The
situation is similar at high magnetic fields, since only the
lowest (spin degenerate) Landau level is partially occupied, hence
the system acts as a metal and external potential is almost
perfectly screened. However, decreasing magnetic field explicitly
implies that the degeneracy $(\sim l^{-2})$ is also reduced
therefore the lowest Landau level can not accommodate all the
electrons. This essentially means that the Fermi energy of the
system lies between two consequent Landau levels all over the
sample, this situation corresponds to Fig.~\ref{fig:fig3}c. Now,
once the higher Landau level is being occupied the $e-e$
interaction forces the system to establish an electrostatic
equilibrium by minimizing the energy. Therefore the IS become
narrower and higher filling fractions are present,
Fig.~\ref{fig:fig3}b. Even lowering the degeneracy, these IS
become more and more narrow until their widths become smaller or
comparable with the quantum mechanical length scales as observed
in Fig~\ref{fig:fig1}.

These results point out clearly that even in a single QHP the edge
states exhibit different behavior by changing their widths.
Moreover, at a very clean sample (where no long-range potential
fluctuations are present) the coherence of the edge state is lost
due to averaging of the phase. On the other hand, if an IS is well
separated and sufficiently wide, it will become ``more'' coherent.
This is true, if the current is confined to these regions, which
we will demonstrate in the next section. We should also remind
that our classical approach in defining the current disables us to
define a phase coherent transport in the sense of LB-ES. Although
there are attempts in defining a coherent transport including
spin-orbit coupling, due to large Zeeman splitting the arguments
of this model~\cite{Tugruls} are rather energetically irrelevant.
However, our hand waving arguments coincides with the findings of
most recent Mach-Zehnder experiments~\cite{Roche07:QHP}, where the
amplitude of the visibility is measured as a function of applied
magnetic field. The reduction of this amplitude is observed at the
higher and the lower edge of the QHP, meanwhile a local maxima
occurred closer to the higher edge.
\section{The local Ohm's law and its implication\label{sec:sec2}}
In this section we briefly summarize the previously developed
local Ohm's law applied to a 2DEG under quantized Hall conditions.
This model assumes that the Ohm's law is valid on the scales
larger than the Fermi wavelength and a local equilibrium is
established. It is essentially based on the argument that if the
local potential and electron distribution is known one can
calculate the local current distribution with in the linear
response regime, i.e. at very small currents.
\begin{figure}
{\centering
\includegraphics[width=.8\linewidth]{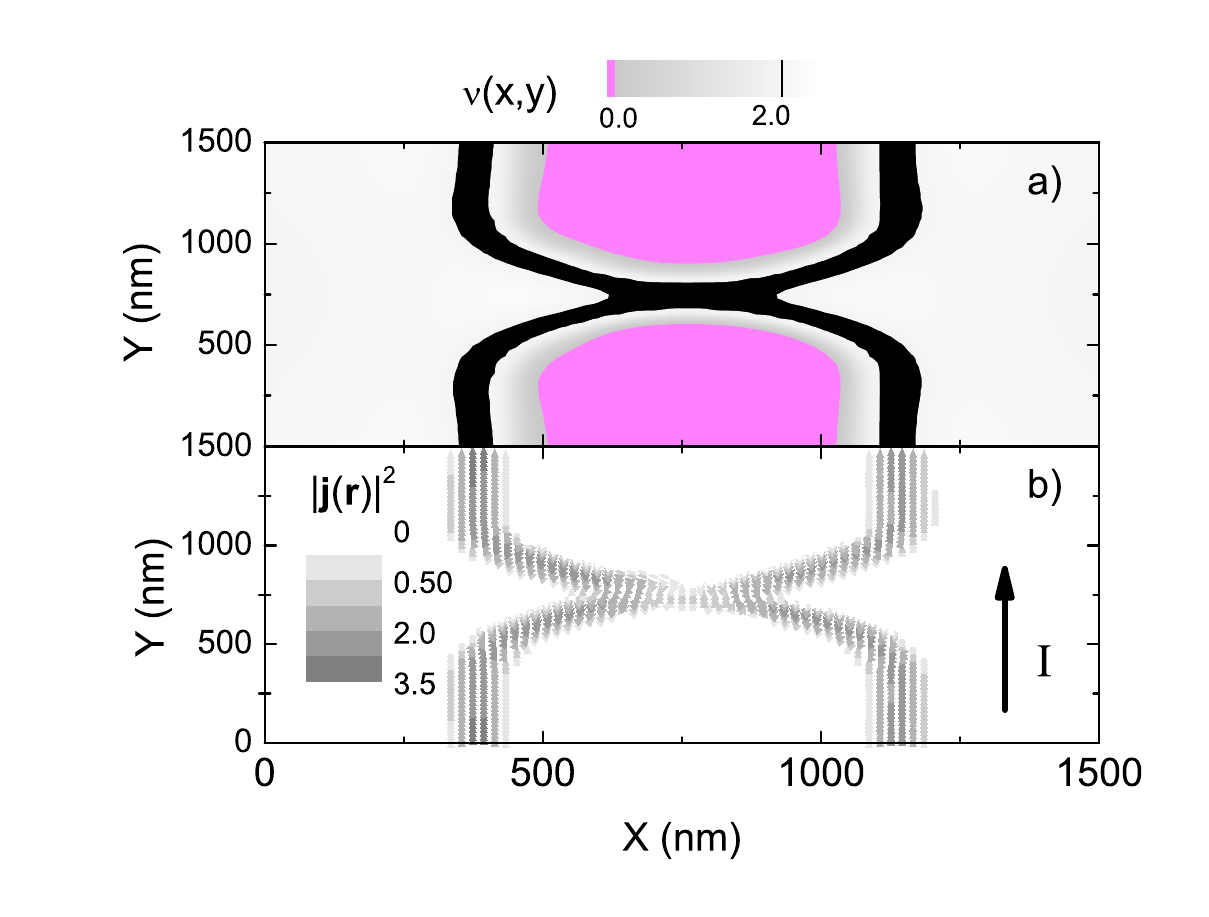}
\caption{ \label{fig:fig4} Gray scale plot of the density (a) and
current distribution within the $\nu=2$ quantized Hall plateau. An
electric field is applied in the $y-$ direction resulting a fixed
current ($j({\bf r})=j_{0}(0,0.1)$), where the 1D current density
is set to $-1.05\times10^{-2}$ A/m. All the other parameters are
the same as Fig.~\ref{fig:fig3} for $B=7.2$ T.}}
\end{figure}
The local Ohm's law assumes that the relation \be
\textbf{E}(\textbf{r})=\hat{\rho}(\textbf{r}).\textbf{j}(\textbf{r})
\ee is valid locally, i.e on the scales of the Fermi wavelength.
The next step is to satisfy the Maxwell equation \be \bf{\nabla}
\times \textbf{E}(\textbf{r})=0 \ee and the equation of continuity
\be \bf{\nabla}\cdot\textbf{j}(\textbf{r})=0\ee simultaneously,
for a stationary situation. From the above equations one can
obtain the local current distribution, if the local resistivity
tensor is known for an external fixed current\be
I=\int_A{\textbf{j}(\textbf{r})d\textbf{r}}.\ee Here, we assume
that $\hat{\rho}(\textbf{r})$ is nothing but the inverse of the
conductivity tensor which is related to the electron distribution
via some reasonable conductivity model. For consistency reasons we
assume a Gaussian broadened spectral function \be
A_n(E)=\frac{e^{(-[\frac{E_n-E}{\Gamma}]^2)}}{\sqrt{\pi}\Gamma},
\ee where $\Gamma$ is the broadening parameter defined by the
impurity potential, which we set $\Gamma/\hbb\omc=0.3$ in order to
smear out the narrow ISs. For the given spectral function the
longitudinal conductivity is calculated as \be
\sigma_l=\frac{2e^2}{h}\int_{-\infty}^{\infty}dE\large[-
\frac{\partial f}{\partial E}\large]\sum_{n=0}^{
\infty}\left(n+\frac{1}{2}\right) [e^{(-[\frac{E_n-E}{\Gamma_{\rm
imp}}]^2)}] \label{eq:sigmaL}\ee whereas the Hall conductivity is
simply the Drude result\be \sigma_H=\frac{2e^2}{h}\nu
\label{eq:sigmaH}.\ee One can already see the essence of the local
model related with the existence of the ISs: the longitudinal
conductivity vanishes if the filling factor is an integer, which
is the case when considering an IS. This is a result of the Landau
gap, if the Fermi energy lies in between two Landau levels there
are no available states at the Fermi level. Therefore, the
conductivity and consequently the longitudinal resistivity
vanishes: \begin{figure} {\centering
\includegraphics[width=.8\linewidth]{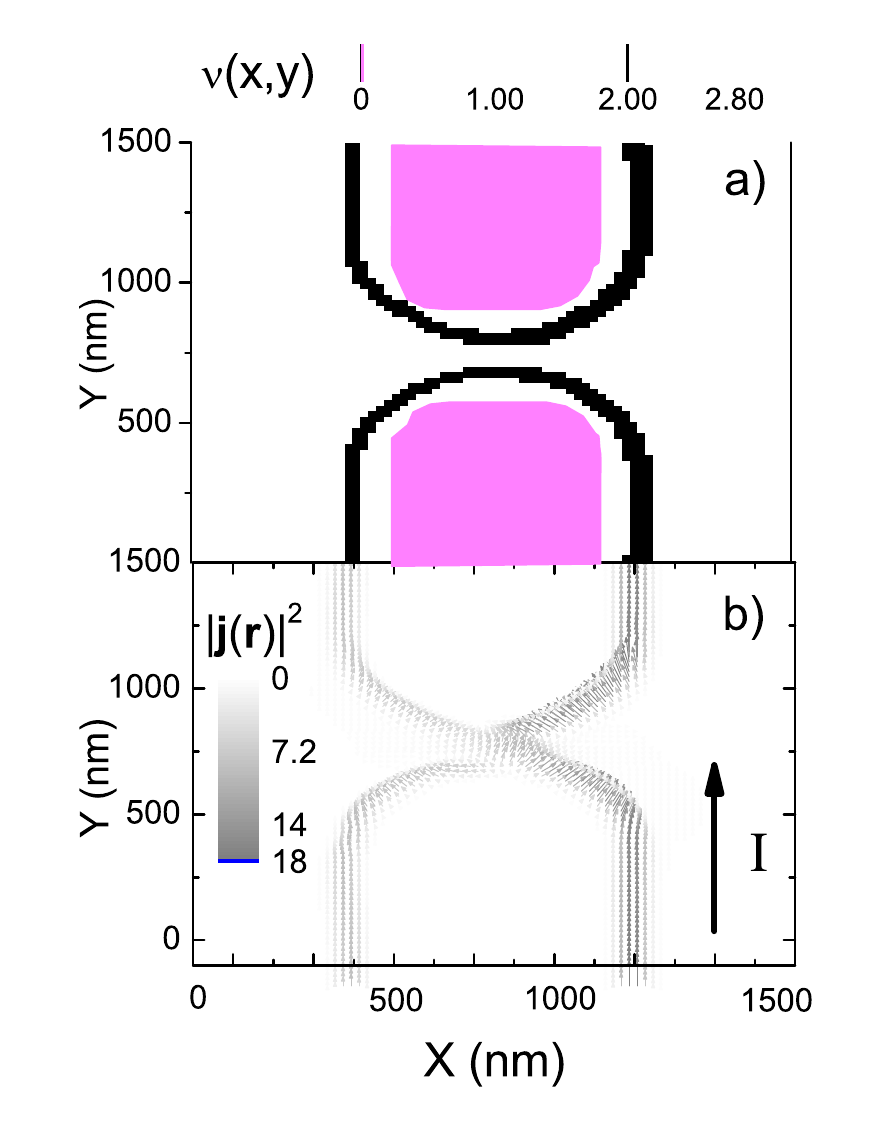}
\caption{ \label{fig:fig5} Gray scale plot of the electron (a) and
current density (b), considering a sufficiently large current
density,  $-4.20\times10^{-2}$ A/m, considering $B=7.2$ T. The
Asymmetry is induced by this current is significant on the
righthand side of the sample.}}
\end{figure}\be
\rho_l(x,y)=\frac{\sigma_l(x,y)}{\sigma^2_l(x,y)+\sigma^2_H(x,y)}
=\frac{0}{0^2+\sigma_H^2}=0.\ee In the linear response regime we
assume that the relation \be
V((x,y);I)-V((x,y);0)\approx\mu^*((x,y);I)-\mu^*_{\rm eq} ,\ee
holds, i.e. the imposed external current does not change the
density and potential profile, thereby the position dependent
electrochemical potential can still be considered as constant. In
Fig.\ref{fig:fig4} we plot the current distribution corresponding
to a density distribution shown in Fig.\ref{fig:fig3}. It is
apparent that the current is confined to the regions where a
well-developed IS is present. Therefore, at least in the linear
response regime, once the spatial distribution of the ISs is
obtained the current distribution can be calculated mutually.
However, when the imposed current is strong enough to influence
the density distribution, one should calculate the position
dependent electrochemical potential $\mu^*(\textbf{r})$ via \be
E(\textbf{r})=\nabla \mu^*(\textbf{r})/e .\ee This relation brings
a new set of self-consistent equations, which have to be solved
numerically. In our (numerical) calculation scheme, we start with
a small total current and in each iteration step increase the
strength of $I$ slowly so that a convergence is guaranteed. The
details of the calculation is given elsewhere~\cite{Sefa08:}.

The density and current distribution is shown in
Fig.\ref{fig:fig5} for a situation where the external current is
sufficiently large. We observe that, the IS on the right side of
the QPC is relatively larger than the one of at the left side. The
physics leading to this effect is rather simple: when an external
current is driven, a Hall potential is built which tilts the
Landau levels, therefore the electrons have to compensate this
potential change. However, one can not add more electrons to the
IS since there are no states available. The only way to screen the
Landau tilting is to increase the widths of the ISs on one side
where the Hall potential is
larger~\cite{Guven03:115327,velocitynldeniz07}. This effect
clearly demonstrates the importance of $e-e$ interactions. Without
the self-consistent treatment one would only get a higher chemical
potential one one side, however, the widths of the edge channels
would remain unchanged. The asymmetry induced by the large
external current on the widths of the IS is also observed in the
local probe experiments, which is a strong evidence that our model
of local Ohm's law is relevant to explain the edge physics of the
IQHE.
\section{Summary}
In conclusion we have demonstrated that the spatial distribution
of the current carrying incompressible strips vary depending on
the magnetic field strength considering a quantum point contact
induced at a 2DEG. Therefore, in interpreting the experimental
results one should keep in mind that the so-called edge states may
present different transport and coherency properties even within a
single quantum Hall plateau. Our electrostatic model relies on the
self-consistent solution of the Poisson and Schr\"odinger
equations using numerical techniques. We have explicitly shown
that the $e-e$ interactions result in the formation of
incompressible and compressible regions. Based on the overlap of
the quantum mechanical wave functions and the local broadening of
the DOS within the incompressible strips, which has an extend
smaller or comparable to the Fermi wavelength, we performed a
spatial average over $\lambda_{\rm F}$ to cure the artifacts of
the TFA. The second set of self-consistent equations has been
introduced to obtain the current distribution, namely solving the
Maxwell equation and equation of continuity simultaneously. In the
small current regime, we have shown that the current is confined
to the incompressible strips since the back scattering is
suppressed and the longitudinal resistivity vanishes. Meanwhile,
the Hall resistivity assumes its quantized value. We have also
shown that, considering sufficiently large external currents
induces an asymmetry both in the density and in the current
distribution, which we think is an interesting issue to check with
the experiments. \ack The author would like to thank R.R.
Gerhardts for his initiation, supervision and contribution in
developing the model, S. Arslan and A. Weichselbaum in providing
the data concerning the solution of the 3D Poisson equation and D.
Eksi and S. Aktas for all their contribution in the improvement of
the numerical code. The enlightening discussions with V. Golovach
is also highly appreciated. This work was supported by the NIM
Area A and SFB 631.
\section*{References}

\end{document}